\documentclass[showpacs,amssymb,aps,twocolumn]{revtex4}
\usepackage{amsmath}
\usepackage{pstricks}
\usepackage{pst-plot}
\usepackage{pst-node}
\usepackage{graphicx}
\usepackage[latin1]{inputenc}
\begin{document}

\title{Thermal Operator and Cutting Rules at Finite Temperature and
  Chemical Potential}

\author{F. T. Brandt$^{a}$, Ashok Das$^{b,c}$, Olivier Espinosa$^{d}$,
  J. Frenkel$^{a}$ and Silvana Perez$^{e}$}
\affiliation{$^{a}$ Instituto de Física, Universidade de São
Paulo, São Paulo, BRAZIL}
\affiliation{$^{b}$ Department of Physics and Astronomy,
University of Rochester,
Rochester, New York 14627-0171, USA}
\affiliation{$^{c}$ Saha Institute of Nuclear Physics, 1/AF Bidhannagar, Calcutta 700064, INDIA}
\affiliation{$^{d}$ Departamento de Física, Universidad
Técnica Federico Santa María, Casilla 110-V, Valparaíso, CHILE}
\affiliation{$^{e}$ Departamento de Física, 
Universidade Federal do Pará, 
Belém, Pará 66075-110, BRAZIL}

\bigskip
%\date{}

\begin{abstract}
In the context of scalar field theories, both real and complex, we
derive the cutting description at finite temperature (with zero/finite
chemical potential) from the cutting rules at zero temperature through
the action of a simple thermal operator. We give an alternative algebraic
proof of the largest time equation which brings out the underlying
physics of such a relation. As an application of the cutting
description, we calculate the imaginary part of the one loop retarded
self-energy at zero/finite temperature and finite chemical potential
and show how this description can be used to calculate the dispersion
relation as well as the full physical self-energy of thermal particles.
\end{abstract}


\maketitle

\section{Introduction}

The imaginary part of a thermal amplitude (with or without a chemical potential) 
has been studied from various points of view by several people both in the imaginary 
time formalism \cite{weldon,jeon} and in the real time formalisms of thermofield dynamics 
\cite{kobes} and closed time path \cite{niegawa,niegawa1,Bedaque:1996af}. In particular, deriving the cutting rules for the imaginary part of an amplitude is of great practical interest. In attempting to derive these  
rules, two important ingredients are involved.
First, one should give a diagrammatic representation to the imaginary part of the amplitude and 
second (which is crucial) one should show that these diagrams allow for a cutting 
description. The earlier attempts at a cutting description of thermal amplitudes, both in the imaginary time formalism \cite{jeon} and in thermofield dynamics \cite{kobes}, succeeded in giving the imaginary part of an amplitude a diagrammatic representation. However, they ran into the difficulty of showing
that these diagrams allow for a cutting description at higher orders beyond one loop. This is primarily due to the fact that at higher orders at finite temperature, graphs with internal isolated islands of
circled/uncircled vertices (circled/uncircled vertices and propagators are explained in section {\bf II}) 
do not vanish individually as a consequence of energy conservation, as they do at zero temperature. A cutting description, on the other hand, requires that such internal isolated islands should not be present in the imaginary part of the amplitude so that the only nonvanishing contributions come from diagrams where circled/uncircled vertices form connected regions. Subsequently,  
it has been shown \cite{Bedaque:1996af} in the closed time path formalism at finite temperature
and zero chemical potential, that such internal isolated islands vanish when summed over the internal thermal indices. As a result,  a cutting description holds for  finite temperature
amplitudes 
much like at zero temperature \cite{niegawa,niegawa1,Bedaque:1996af,das:book97,Landshoff:1996ta}. 
There are, however, two
important differences at finite temperature in comparison with the
zero temperature analysis \cite{cutkosky,tHooft:1973jh}.   
First, at finite temperature, there is a
doubling of fields (we follow the notations and conventions in 
\cite{das:book97} and
call these fields $\pm$). Second, unlike at zero temperature where a
cutting description holds graph by graph, at finite temperature such a
description holds only when we sum over classes of graphs involving
intermediate ``$\pm$" vertices. The proof of such a cutting
description, even at zero chemical potential at finite temperature, however,
turns out to be rather involved \cite{Bedaque:1996af,das:book97}. 

On the other hand, in a series of papers \cite{Espinosa:2003af,Espinosa:2005gq,silvana1,Brandt:2006rv}
we have shown recently that both in
the real time as well as in the imaginary time 
formalisms \cite{das:book97,kapusta:book89,lebellac:book96}, 
a finite
temperature Feynman graph can be represented as a simple
multiplicative thermal operator acting on the corresponding zero
temperature Feynman graph (in the imaginary time formalism the zero
temperature graph is that of the corresponding Euclidean field
theory). Such a thermal operator representation holds even when the
chemical potential is nonzero although in this case the thermal operator is a bit
more complicated \cite{Inui:2006jf,Brandt:2006vy}. The thermal
operator has many nice properties
including the fact that it is real and the  proof of such a thermal
operator representation of Feynman graphs is most direct in a mixed
space (although the thermal operator representation holds even in
momentum space). 

While the thermal operator representation is useful from various
points of view, since the thermal operator is real, a natural
application of this relation will be in deriving the cutting rules at
finite temperature from those at zero temperature. This may
simplify the proof of the cutting description given in
\cite{Bedaque:1996af,das:book97}
and may 
allow us to extend a proof of the cutting description for theories at
finite temperature and chemical potential. However, in trying to derive the
finite temperature cutting description from that at zero temperature,
one faces two difficulties. First, at zero temperature classes
of diagrams - those containing isolated circled/uncircled internal
vertices- vanish by energy conservation and it is known that the
thermal operator acting on such terms can lead to nonvanishing
contributions at finite temperature. Therefore, it is not clear how to
obtain a proof for the cutting description of a general graph at any
arbitrary loop. The second difficulty is associated with the fact that
at finite temperature, there is a doubling of fields which is not
present at zero temperature. The second difficulty is easier to
handle, as we have indicated in our earlier papers. Namely, we simply take
the zero temperature theory to correspond to the theory with the
doubled degrees of freedom obtained from finite temperature by taking
the zero temperature limit. Of course, the additional degrees of
freedom (in our notation the ``$-$" fields) lead to vanishing
contribution at zero temperature. However, as we will see, these
additional vertices are quite essential in giving a simpler derivation
of the cutting description at finite temperature. With this, the first
difficulty may be avoided if classes of diagrams containing isolated
circled/uncircled internal vertices add up to zero without the use of
explicit energy conservation. As we will show, this is exactly what happens
which leads to a simpler derivation of the cutting description at
finite temperature with and without a chemical potential. 

The paper is organised as follows. In section {\bf II}, we
systematically study a real scalar field theory (without a chemical
potential) with doubled degrees of freedom at zero temperature and
derive the cutting description for such a theory which then leads naturally  to
the cutting description at finite temperature through the application
of the thermal operator. In this section, we  give an alternative 
algebraic derivation of the largest/smallest time equation which also
brings out the physical meaning associated with these. We derive
various identities associated with the relevant Green's functions (and
their physical origin) which are quite crucial in proving the cutting
description. We also work out  in this approach the imaginary part of
the one loop retarded self-energy in the $\phi^{3}$ theory.  In
section {\bf III}, we 
extend all of our analysis of section {\bf II} and derive the cutting
description for a complex scalar field (interacting with a real scalar
field) at finite temperature and nonzero chemical 
potential. Such an
analysis has not been carried out earlier and our analysis shows that
the proof is no more difficult than in the case of a vanishing
chemical potential. As an example, we calculate the imaginary part of
the one loop retarded self-energy for the real scalar field
interacting with a complex scalar  field. Furthermore, we show how one
can use the circled diagrams to 
calculate the full retarded self-energy (and not just the imaginary
part) as well as the dispersion relations at any temperature. We
conclude  with a brief summary in section {\bf IV}. In 
appendix {\bf A}, we give a brief derivation of an identity used in
deriving the largest time equation. This derivation also shows that
the largest time equation holds for any theory (with $n$-point
interactions) and not just the $\phi^{3}$ theory that we deal with in
the paper for simplicity. In appendix {\bf B}, we give the spectral
decomposition  for the components of the
scalar propagator at finite temperature and chemical potential.

\section{Cutting Rules at Finite $T$ for $\mu = 0$}

In this section, we will study the cutting rules for the $\phi^{3}$ theory at
finite temperature with a vanishing chemical potential through the
thermal operator representation. The cutting description for such a
theory  has been
studied earlier in detail \cite{Bedaque:1996af,das:book97} where the
derivation of the rules were quite 
nontrivial and appeared to be quite distinct from those at zero
temperature. From the point of view of the thermal operator
representation, we will see that the derivation of these rules at
finite temperature are completely parallel to those at zero
temperature. We consider a real scalar field with
a cubic interaction for simplicity for which the $2\times 2$ matrix
propagators in the closed time path formalism are well known both in
the momentum space representation as well as in the mixed space
representation. Defining the propagator matrix as 
\begin{equation}
\Delta^{(T)} = \left(\begin{array}{cc}
\Delta_{++}^{(T)} & \Delta_{+-}^{(T)}\\
 & \\
\Delta_{-+}^{(T)} &\Delta_{--}^{(T)}
\end{array}\right),
\end{equation}
we note that in the momentum space the components have the explicit forms
\begin{eqnarray}
\Delta_{++}^{(T)} (p) & = & \lim_{\epsilon\rightarrow 0}\
\frac{i}{p^{2}-m^{2} + i\epsilon} + 2\pi
n(|p_{0}|) \delta (p^{2}-m^{2}),\nonumber\\
\Delta_{+-}^{(T)} (p) & = & 2\pi \left(\theta(-p_{0}) +
n(|p_{0}|)\right) \delta (p^{2}-m^{2}),\nonumber\\
\Delta_{-+}^{(T)} (p) & = & 2\pi \left(\theta(p_{0}) +
  n(|p_{0}|)\right)\delta (p^{2}-m^{2}),\label{Tmomprop}\\
\Delta_{--}^{(T)} (p) & = & \lim_{\epsilon\rightarrow 0}\  -
\frac{i}{p^{2}-m^{2}-i\epsilon} + 2\pi n(|p_{0}|) \delta
(p^{2}-m^{2}),\nonumber
\end{eqnarray}
with $n(|p_{0}|)$ denoting the bosonic distribution function, while in
the mixed space they can be written as 
\begin{widetext}
\begin{eqnarray}
\Delta_{++}^{(T)} (t,E) & = & 
\frac{1}{2E}\left[\theta(t) e^{-iEt} +
\theta(-t)e^{iEt} + n(E)(e^{-iEt} +
e^{iEt})\right],\nonumber\\
\Delta_{+-}^{(T)} (t,E) & = & \frac{1}{2E}\left[n(E) e^{-iEt} + (1 +
n(E)) e^{iEt}\right],\nonumber\\
\Delta_{-+}^{(T)} (t,E) & = & \frac{1}{2E}\left[(1 + n(E)) e^{-iEt} +
n(E) e^{iEt}\right],\nonumber\\
\Delta_{--}^{(T)} (t,E) & = & 
\frac{1}{2E}\left[\theta(t) e^{iEt} + \theta (-t)
e^{-iEt} + n(E) (e^{-iEt} +
e^{iEt})\right],\label{Tprop}
\end{eqnarray}
\end{widetext}
where $E = \sqrt{\vec{p}^{\ 2} + m^{2}}$ and we have dropped the
$i\epsilon$ factors in the exponent for simplicity. For 
future use, we note from the structures of the propagators in (\ref{Tprop}) that the 
KMS condition (periodicity condition) \cite{kubo,martin} at finite 
temperature leads in the mixed space to the relation
\begin{equation}
\Delta_{-+}^{(T)} (t,E) =  \Delta_{+-}^{(T)} (t + i\beta,E),\label{KMS}
\end{equation}
where $\beta$ denotes the inverse temperature in units of the Boltzmann constant.

At zero temperature, the components of the propagators take the
respective forms 
\begin{eqnarray}
\Delta_{++}^{(T=0)} (p) & = & \lim_{\epsilon\rightarrow 0}\
\frac{i}{p^{2}-m^{2}+i\epsilon},\nonumber\\
\Delta_{+-}^{(T=0)} (p) & = & 2\pi \theta (-p_{0}) \delta
(p^{2}-m^{2}),\nonumber\\
\Delta_{-+}^{(T=0)} (p) & = & 2\pi \theta (p_{0}) \delta
(p^{2}-m^{2}),\nonumber\\
\Delta_{--}^{(T=0)} (p) & = & \lim_{\epsilon\rightarrow 0}\ -
\frac{i}{p^{2}-m^{2}-i\epsilon},\label{zeroTmomprop}
\end{eqnarray}
and 
\begin{eqnarray}
\Delta_{++}^{(T=0)} (t,E) & = & 
\frac{1}{2E}\left[\theta (t) e^{-iEt} + \theta (-t)
e^{iEt}\right],\nonumber\\
\Delta_{+-}^{(T=0)} (t,E) & = & \frac{1}{2E}\ e^{iEt},\nonumber\\
\Delta_{-+}^{(T=0)} (t,E) & = & \frac{1}{2E}\ e^{-iEt},\nonumber\\
\Delta_{--}^{(T=0)} (t,E) & = & 
\frac{1}{2E}\left[\theta (t) e^{iEt} + \theta (-t)
e^{-iEt}\right]. \label{zeroTprop}
\end{eqnarray}
As we have noted in earlier 
papers \cite{silvana1,Brandt:2006rv}, 
the finite temperature
propagator in the mixed space (\ref{Tprop}) is easily seen to be
related to that at zero temperature (\ref{zeroTprop}) through the
thermal operator as 
\begin{equation}
\Delta_{ab}^{(T)} (t, E) = {\cal O}^{(T)}(E)\Delta_{ab}^{(T=0)}
(t,E), \quad a,b =\pm, \label{factorization} 
\end{equation}
where the basic thermal operator is defined to be
\begin{equation}
{\cal O}^{(T)} (E) = 1 + n(E) (1-S(E)),\label{tor}
\end{equation}
with $S(E)$ representing a reflection operator that changes
$E\rightarrow -E$. 
The important thing to note here is that the basic thermal operator is
independent of the time coordinates and as a result any Feynman graph
at finite temperature factorizes leading to a thermal operator
representation for the graph \cite{silvana1,Brandt:2006rv}.

Let us study the theory at zero temperature with doubled degrees of
freedom resulting from the zero temperature limit of the finite
temperature theory in the closed time path formalism. We know that at
zero temperature, the ``$+$" components of the fields define the
dynamical variables and there is no contribution of the ``$-$"
components of the fields to the amplitudes of the dynamical
fields. So, in some sense adding these extra components at zero
temperature correspond to adding vanishing
contributions. Nevertheless, we will see that with these added
(vanishing) contributions at zero temperature, the proof of the
cutting rules at finite temperature become completely parallel to that at zero temperature
 through the thermal operator
representation. We would like to emphasize here that if one were to
calculate physical Green's functions such as the retarded and advanced
propagators at zero temperature, such a doubling of degrees of freedom
is inevitable. 

In order to determine the imaginary part of a Feynman
amplitude diagrammatically, we need to introduce a diagrammatic
representation for the complex conjugate of a graph. This can be done
in the standard manner by enlarging the theory with circled vertices and
propagators in the following way. We note that the propagators are
time ordered Green's functions and as usual, we can decompose them
into their positive and negative frequency components as 
\begin{eqnarray}
\Delta_{ab}^{(T=0)} (t,E) & = & \theta (t) \Delta_{ab}^{(T=0) (+)} (t,E)\nonumber\\
 & & +  \theta(-t) \Delta_{ab}^{(T=0) (-)} (t,E).\label{frequencydecomp}
\end{eqnarray}
The positive and the negative frequency parts of the propagators at
zero temperature can be read out from (\ref{zeroTprop}).
Furthermore, given this decomposition, we can define a set of circled
propagators as 
\begin{widetext}
\begin{eqnarray}
\hbox{\pspicture(0,-.075)(3,1)
      \psline(.5,0)(2.5,0)
      \rput[t]{0}(.5,-.2){$a,t_{1}$}
      \rput[t]{0}(2.3,-.2){$b,t_{2}$}
      \endpspicture} & = & \Delta_{ab}^{(T=0)}(t_{1}-t_{2}, E) =
    \theta(t_{1}-t_{2}) \Delta_{ab}^{(T=0)(+)} (t_{1}-t_{2},E) + \theta(t_{2}-t_{1})\Delta_{ab}^{(T=0)(-)}(t_{1}-t_{2},E)\nonumber\\
\hbox{\pspicture(0,-.075)(3,1)
      \psline(.5,0)(2.5,0)
      \pscircle(2.5,0){.14}
      \rput[t]{0}(.5,-.2){$a,t_{1}$}
      \rput[t]{0}(2.3,-.2){$b,t_{2}$}
      \endpspicture} & = & \Delta_{ab}^{(T=0)} (t_{1}-\underline{t_{2}},E) = \Delta_{ab}^{(T=0)(-)}(t_{1}-t_{2},E)\nonumber\\
\hbox{\pspicture(0,-.075)(3,1)
      \psline(.5,0)(2.5,0)
      \pscircle(.5,0){.14}
      \rput[t]{0}(.5,-.2){$a,t_{1}$}
      \rput[t]{0}(2.3,-.2){$b,t_{2}$}
      \endpspicture} & = & \Delta_{ab}^{(T=0)}(\underline{t_{1}}-t_{2},E) = \Delta_{ab}^{(T=0)(+)}(t_{1}-t_{2},E)\nonumber\\
\hbox{\pspicture(0,-.075)(3,1)
      \psline(.5,0)(2.5,0)
      \pscircle(.5,0){.14}
      \pscircle(2.5,0){.14}
      \rput[t]{0}(.5,-.2){$a,t_{1}$}
      \rput[t]{0}(2.3,-.2){$b,t_{2}$}
      \endpspicture} & = &
    \Delta_{ab}^{(T=0)}(\underline{t_{1}}-\underline{t_{2}},E) =
    \theta(t_{1}-t_{2})\Delta_{ab}^{(T=0)(-)}(t_{1}-t_{2},E)+
    \theta(t_{2}-t_{1})\Delta_{ab}^{(T=0)(+)}(t_{1}-t_{2},E). \label{circledprop}
\end{eqnarray}
\smallskip

%\centerline{Fig. 1: Circled propagators}
\medskip

\end{widetext}
Here $a,b=\pm$ and we have followed the standard convention that an
underlined coordinate denotes a vertex that is circled. We note that
the propagator with both ends circled is the anti-time ordered
propagator which is easily seen to be the complex conjugate of the
original propagator in the momentum space. (We note here that for the
``$++$''and ``$--$''components, the anti-time ordered propagators are
the complex conjugates of the time ordered propagators even in the
mixed space. However, for the
``$+-$''and the ``$-+$''components this is not true as they are not
even functions of momenta in the momentum space.) Similarly, we
introduce a circled interaction vertex to be the complex conjugate of
the original vertex. For real coupling constants, this corresponds to
simply changing the sign of the interaction vertex. 
\begin{eqnarray}
\hbox{\pspicture(0,-.075)(3,1.5)
      \psline(1,0.5)(2,0)
      \psline(1,-0.5)(2,0)
      \psline(2,0)(3,0)
      \psdots(2,0)
      \rput[b]{0}(2.4,.1){$a$}
      \rput[l]{0}(1.7,.3){$a$}
      \rput[l]{0}(1.7,-.3){$a$}
      \endpspicture}
 &=& -i(a) g,\nonumber\\
\hbox{\pspicture(0,-.075)(3,1.5)
      \psline(1,0.5)(2,0)
      \psline(1,-0.5)(2,0)
      \psline(2,0)(3,0)
      \psdots(2,0)
      \pscircle(2,0){.14}
      \rput[b]{0}(2.4,.1){$a$}
      \rput[l]{0}(1.7,.3){$a$}
      \rput[l]{0}(1.7,-.3){$a$}
      \endpspicture}
 &=& i(a) g,\label{circledvertices}
\end{eqnarray}
\smallskip

%\centerline{Fig. 2: Circled vertices}
\smallskip

\noindent where $a=\pm$. With these it is clear that a graph with all
vertices circled
is the complex conjugate of the corresponding graph of the original
theory (graph with no vertex circled) in the momentum space and
consequently the imaginary
part of a graph of the original theory in momentum space can be given
a diagrammatic representation.  

Let us note here that one can also decompose the finite temperature
propagators in (\ref{Tprop}) into positive and negative frequency
parts through the definition (\ref{frequencydecomp}) (we note here that at finite temperature each of these functions contains both positive as well as negative frequency components, but the nomenclature carries over from zero temperature) and correspondingly define a set of circled propagators at
nonzero temperature. It is clear from their forms that all such
propagators are related to the zero temperature propagators by the
same basic thermal operator in (\ref{tor}), namely, 
\begin{eqnarray}
\Delta_{ab}^{(T)} (t_{1}-t_{2},E) & = & {\cal O}^{(T)}(E) \Delta_{ab}^{(T=0)} (t_{1}-t_{2},E),\nonumber\\
\Delta_{ab}^{(T)} (t_{1}-\underline{t_{2}},E) & = & {\cal O}^{(T)} (E) \Delta_{ab}^{(T=0)} (t_{1}-\underline{t_{2}},E),\nonumber\\
\Delta_{ab}^{(T)} (\underline{t_{1}}-t_{2},E) & = & {\cal O}^{(T)} (E) \Delta_{ab}^{(T=0)} (\underline{t_{1}}-t_{2},E),\nonumber\\
\Delta_{ab}^{(T)} (\underline{t_{1}}-\underline{t_{2}},E) & = & {\cal O}^{(T)} (E) \Delta_{ab}^{(T=0)} (\underline{t_{1}}-\underline{t_{2}},E).\label{circled}
\end{eqnarray}
Such a relation, in turn, allows us to relate the imaginary part of a
Feynman graph at finite temperature to that at zero temperature
through the thermal operator.  (We recall that interactions vertices
are not modified at finite temperature and that the thermal operator
is real.)  

From the definition of the circled propagators, we can derive an
interesting relation  conventionally known as the largest/smallest
time equation. Here we give an algebraic derivation of this relation
that also brings out the physics underlying such a relation. Let us
note from the definition of the circled propagators (\ref{circledprop}) that 
\begin{widetext}
\begin{eqnarray}
\Delta_{ab}^{(T=0)} (t_{1}-t_{2},E) - \Delta_{ab}^{(T=0)}(\underline{t_{1}}-t_{2},E) & = &  \Delta_{ab}^{(T=0)} (t_{1}-\underline{t_{2}},E) - \Delta_{ab}^{(T=0)} (\underline{t_{1}}-\underline{t_{2}},E)\nonumber\\
& = & \theta (t_{2}-t_{1})\left(\Delta_{ab}^{(T=0)(-)} (t_{1}-t_{2},E) - \Delta_{ab}^{(T=0)(+)} (t_{1}-t_{2},E)\right).\label{lte}
\end{eqnarray}
\end{widetext}
Namely, adding to a propagator with an uncircled vertex another with
the corresponding vertex circled makes the time coordinate of this
vertex advanced with respect to the second time coordinate. As a
result, it follows that 
\begin{eqnarray}
\theta(t_{1}-t_{2})(\Delta_{ab}^{(T=0)} (t_{1}-t_{2},E) - \Delta_{ab}^{(T=0)} (\underline{t_{1}}-t_{2},E)) & = & 0,\nonumber\\
\theta(t_{1}-t_{2})(\Delta_{ab}^{(T=0)} (t_{1}-\underline{t_{2}},E) -
\Delta_{ab}^{(T=0)} (\underline{t_{1}}-\underline{t_{2}},E)) & = &
0.\nonumber\\
& & \label{lte1}
\end{eqnarray}
These relations are really at the heart of the largest time equation
and from the above equations we note that the vanishing of the
relations in (\ref{lte1}) holds independent of whether the second vertex is circled or
not. Thus, for simplicity, in this $\phi^{3}$ theory let us consider
the sum of the  two diagrams shown in figure \eqref{fe1}
with vertices $(t_{i},a_{i}), i=1,2,3$ uncircled.
\begin{widetext}
\begin{center}
\begin{figure}[ht!]
\includegraphics[scale=0.4]{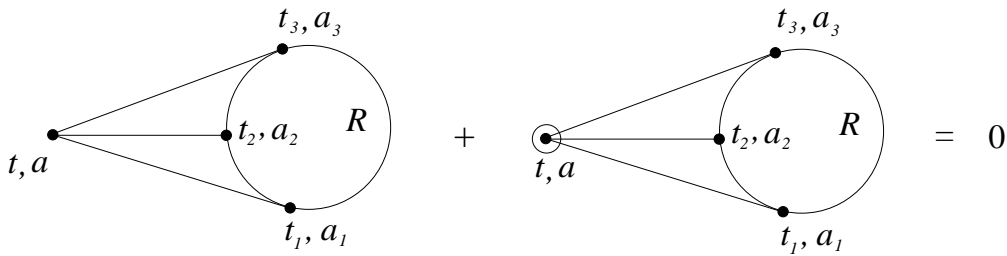}
\caption{Largest time equation.}\label{fe1}
\end{figure}
\end{center}
\end{widetext}

\noindent where $R = R (a_{1},a_{3})$ stands for the rest of the graph (we will use the notation $R$ through out the paper, but its dependence on the thermal indices $a_{1},a_{3}$ is to be understood). If we assume $t$
to be the largest time (namely, $t>t_{i}, i=1,2,3$), then the sum of the graphs
in figure \eqref{fe1} can be written as (recall that a circled vertex
has an additional negative sign)
\begin{widetext}
\begin{eqnarray}
\lefteqn{\Gamma = R\left[\prod_{i=1}^{3}\theta(t-t_{i})\Delta_{aa_{i}}^{(T=0)}(t-t_{i},E_{i}) -
\prod_{i=1}^{3}
\theta(t-t_{i})\Delta_{aa_{i}}^{(T=0)}(\underline{t}-t_{i},E_{i})\right]}\nonumber\\
 & = &  \frac{R}{4}\left[\sum_{i=1}^{3} 
 \theta(t-t_{i})\left(\Delta_{aa_{i}}^{(T=0)}
   (t-t_{i},E_{i})-\Delta_{aa_{i}}^{(T=0)}(\underline{t}-t_{i},E_{i})\right)\prod_{j\neq i} \theta(t-t_{j})\left(\Delta_{aa_{j}}^{(T=0)}(t-t_{j},E_{j}) + \Delta_{aa_{j}}^{(T=0)}(\underline{t}-t_{j},E_{j})\right)\right.\nonumber\\
   & & \quad \left.+ \prod_{i=1}^{3}\theta(t-t_{i})\left(\Delta_{aa_{i}}^{(T=0)}
   (t-t_{i},E_{i})-\Delta_{aa_{i}}^{(T=0)}(\underline{t}-t_{i},E_{i})\right)\right]
 =  0,\label{lte2}
\end{eqnarray}
\end{widetext}
which follows from (\ref{lte1}). This is known as the largest time
equation, namely, if we take a Feynman graph with the largest time
vertex uncircled and add to it the same graph where the largest time
vertex is circled, then the sum vanishes. (The largest time equation
holds for any theory and a short derivation of the necessary
identities in the case of a theory with $n$-point interactions is
given in appendix {\bf A}.) Physically, this can be seen
from the relation (\ref{lte}) according to which summing over the two
diagrams in the above would make the time coordinate $t$ to be
advanced with respect to at least one of the $t_{i}$'s. On the other
hand, this is not possible if $t$ happens to be the largest time and,
consequently, the sum must vanish. In a similar manner, one can also
derive the smallest time equation that we will not go into. We also
note here that we can also derive the largest/smallest time equation
where a circled vertex is replaced by a ``$-$''vertex, but we do not
go into that for simplicity. (We remark
here parenthetically that if $t$ denotes the largest time, namely,
$t>t_{i}, i=1,2,3,$ then using (\ref{lte1}) we note that
\begin{equation}
\theta (t-t_{i}) \Delta_{aa_{i}}^{(T=0)} (t-t_{i},E_{i}) =
\theta(t-t_{i})\Delta_{aa_{i}}^{(T=0)}(\underline{t}-t_{i},E_{i}),
\end{equation}
and the two factors in the first line of (\ref{lte2}) cancel
identically. This is the most direct way of deriving the largest time
equation in any theory. However, the above derivation clarifies the
underlying physics of such a relation.)

From (\ref{circled}) we see that the circled propagators at finite
temperature are related to those at zero temperature through the basic
thermal operator which is independent of time. It follows, therefore,
that the largest time equation also holds at finite temperature as has
been shown explicitly in \cite{das:book97}. A consequence of the largest time
equation (both at zero as well as at finite temperature) is that if we
take any diagram with all possible circlings of the vertices, then the
sum of all such diagrams must vanish. This follows from the fact that
for any given time ordering, the sum would involve pairs of diagrams with the
largest time vertex uncircled and circled which will cancel
pairwise. Thus, denoting a graph with $n$ vertices by $F^{(T=0)}_{a_{1}\cdots
  a_{n}} (t_{1},\cdots ,t_{n})$ (where we are suppressing the energy
dependence), we have
\begin{equation}
\sum_{\rm circlings} F^{(T=0)}_{a_{1}\cdots a_{n}} (t_{1},\cdots, t_{n}) = 0.
\end{equation}
This, in turn implies that
\begin{eqnarray}
& & F_{a_{1}\cdots a_{n}}^{(T=0)} (t_{1},\cdots, t_{n}) +
F_{a_{1}\cdots a_{n}}^{(T=0)}
(\underline{t_{1}},\cdots,\underline{t_{n}})\nonumber\\
& &\quad = - \sideset{}{^\prime}\sum_{\rm circlings} F_{a_{1}\cdots
  a_{n}}^{(T=0)}(t_{1},\cdots,t_{n}),\label{imaginary}
\end{eqnarray}
where the prime refers to the sum of all circlings except where no
vertices/all vertices are circled. We note that the left hand side is
two times the imaginary part of the diagram (up to a factor of ``$i$'')
in  the momentum space.

It is worth remembering that the internal vertices in a diagram are,
of course, integrated over the respective time coordinates, but in
addition the ``thermal" indices of the internal vertices also need to
be summed over. As a result, the number of diagrams on the right hand
side of (\ref{imaginary}) is, in general, very large. However, a lot
of them vanish and to determine the nontrivial ones that contribute to
the imaginary part of the diagram, we make use of some interesting
identities involving the circled propagators. By direct inspection of
the propagators in (\ref{zeroTprop}) and their positive and negative
frequency parts, it can be easily seen that when only one of the ends
of the propagator is circled, it takes a very simple form, namely, 
\begin{eqnarray}
\Delta_{ab}^{(T=0)} (t_{1}-\underline{t_{2}},E) & = & \Delta_{a\ -a}^{(T=0)} (t_{1}-t_{2},E),\nonumber\\
\Delta_{ab}^{(T=0)} (\underline{t_{1}}-t_{2},E) & = & \Delta_{-b\ b}^{(T=0)} (t_{1}-t_{2},E).\label{identities}
\end{eqnarray}
Here we have introduced the notation $-a = \mp$ for $a=\pm$. As we
will see, these relations are quite crucial in obtaining a cutting
description of graphs. Basically, they say that when one of the ends
of a propagator is circled, the propagator is independent of the
``thermal" index of the circled end. While this may seem  surprising,
it is easy to see that this has a physical origin. Let us recall that
the retarded propagator of the theory is given by (this can be
obtained from the zero temperature limit of the definition in \cite{das:book97}) 
\begin{eqnarray}
& & \Delta_{\rm R}^{(T=0)}(t_{1}-t_{2},E)\nonumber\\
&  & =  \Delta_{++}^{(T=0)}(t_{1}-t_{2},E) - \Delta_{+-}^{(T=0)} (t_{1}-t_{2},E)\nonumber\\
& & = \Delta_{-+}^{(T=0)} (t_{1}-t_{2},E) - \Delta_{--}^{(T=0)} (t_{1}-t_{2},E).
\end{eqnarray}
Putting in the positive and the negative frequency decompositions, the
two relations lead to 
\begin{widetext}
\begin{eqnarray}
\Delta_{\rm R}^{(T=0)}(t_{1}-t_{2},E) & = & \theta(t_{1}-t_{2})\left(\Delta_{++}^{(T=0)(+)}(t_{1}-t_{2},E)-\Delta_{+-}^{(T=0)(+)} (t_{1}-t_{2},E)\right)\nonumber\\
& &\quad + \theta(t_{2}-t_{1})\left(\Delta_{++}^{(T=0)(-)}(t_{1}-t_{2},E) - \Delta_{+-}^{(T=0)(-)}(t_{1}-t_{2},E)\right),\nonumber\\
\Delta_{\rm R}^{(T=0)}(t_{1}-t_{2},E) & = & \theta(t_{1}-t_{2})\left(\Delta_{-+}^{(T=0)(+)}(t_{1}-t_{2},E)-\Delta_{--}^{(T=0)(+)} (t_{1}-t_{2},E)\right)\nonumber\\
& & \quad + \theta(t_{2}-t_{1})\left(\Delta_{-+}^{(T=0)(-)}(t_{1}-t_{2},E) - \Delta_{--}^{(T=0)(-)}(t_{1}-t_{2},E)\right).
\end{eqnarray}
\end{widetext}
On the other hand, by definition the retarded propagator is
proportional to $\theta(t_{1}-t_{2})$ and, therefore, we must have 
\begin{eqnarray}
\Delta_{++}^{(T=0)(-)} & = & \Delta_{+-}^{(T=0)(-)} = \Delta_{+-}^{(T=0)},\nonumber\\
\Delta_{-+}^{(T=0)(-)} &=& \Delta_{--}^{(T=0)(-)} = \Delta_{-+}^{(T=0)},
\end{eqnarray}
where we have used the fact that the positive and the negative
frequency components of $\Delta_{\pm\mp}$ coincide with the respective
propagators. This, in turn, implies that 
\begin{equation}
\Delta_{ab}^{(T=0)(-)} (t_{1}-t_{2},E) = \Delta_{a\ -a}^{(T=0)}(t_{1}-t_{2},E).
\end{equation}
On the other hand, by definition
\begin{equation}
\Delta_{ab}^{(T=0)(-)}(t_{1}-t_{2},E) = \Delta_{ab}^{(T=0)}(t_{1}-\underline{t_{2}},E),
\end{equation}
so that the first of (\ref{identities}) follows. Similarly, looking at
the advanced propagator we can show that
\begin{eqnarray}
\Delta_{ab}^{(T=0)(+)} (t_{1}-t_{2},E) & = & \Delta_{-b\
b}^{(T=0)}(t_{1}-t_{2},E)\nonumber\\
 & = & \Delta_{ab}^{(T=0)} (\underline{t_{1}}-t_{2},E),
\end{eqnarray}
which leads to the second of the relations in (\ref{identities}). It
is clear, therefore, that $\Delta_{\pm\mp}^{(T=0)}$ represent the two
basic propagators at zero temperature (and through the thermal
operator at finite temperature as well). In fact, we note that we can
even express the uncircled and doubly circled propagators in terms of
them as
\begin{eqnarray}
\Delta_{ab}^{(T=0)}(t_{1}-t_{2},E) & = &
\theta(t_{1}-t_{2})\Delta_{-b\ b}^{(T=0)}(t_{1}-t_{2},E)\nonumber\\
& & +
\theta(t_{2}-t_{1})\Delta_{a\ -a}^{(T=0)}(t_{1}-t_{2},E),\nonumber\\
\Delta_{ab}^{(T=0)}(\underline{t_{1}}-\underline{t_{2}},E) & = &
\theta(t_{1}-t_{2})\Delta_{a\ -a}^{(T=0)}(t_{1}-t_{2},E)\nonumber\\
& & + \theta(t_{2}-t_{1})\Delta_{-b\ b}^{(T=0)}(t_{1}-t_{2},E).\label{basic}
\end{eqnarray}
Therefore, $\Delta_{\pm\mp}^{(T=0)}$ truly denote the two basic
propagators in terms of which all other propagators
(circled/uncircled) can be expressed. For completeness, let us note
from (\ref{zeroTprop}) that we can compactly write
\begin{equation}
\Delta_{a\ -a}^{(T=0)} (t_{1}-t_{2},E) = \frac{1}{2E} e^{i(a)
  E(t_{1}-t_{2})}.\label{aminusa}
\end{equation}  
Let us also note some other interesting identities involving the
circled propagators. From the forms of various
propagators, we can easily verify that
\begin{eqnarray}
\sum_{a,b=\pm} (a)(b) \Delta_{ab}^{(T=0)} (t_{1}-t_{2}, E) & = & 0,\nonumber\\
\sum_{a,b=\pm} (a) (b) \Delta_{ab}^{(T=0)} (t_{1}-\underline{t_{2}}, E) & = &
0,\nonumber\\
\sum_{a,b=\pm} (a) (b) \Delta_{ab}^{(T=0)} (\underline{t_{1}}-t_{1}, E) & = &
0,\nonumber\\
\sum_{a,b=\pm} (a) (b) \Delta_{ab}^{(T=0)} (\underline{t_{1}}-\underline{t_{2}}, E)
& = & 0.\label{identity1}
\end{eqnarray}

\begin{figure}[ht!]
\begin{center}
\includegraphics[scale=0.4]{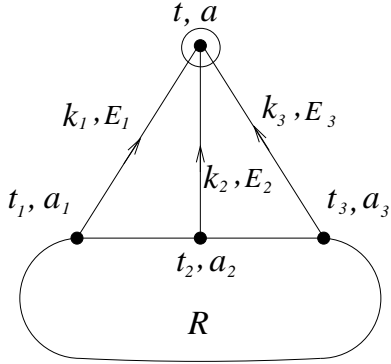}
\end{center}
\caption{A typical diagram with a single isolated internal circled vertex.}\label{fe2}
\end{figure}

With these relations, we are now ready to derive the cutting
description for this theory at zero temperature with doubled degrees
of freedom. First, let us note that in this theory, if we have a
diagram with an isolated internal circled vertex, then its
contribution identically vanishes. In fact, for any value of the
``thermal''index $a$  the contribution (we are
ignoring combinatoric factors as well as the coupling constants) in
the integrand corresponding to the diagram shown in figure \eqref{fe2}.

\noindent has the form (we are suppressing the energy dependence of
the propagators for simplicity)
\begin{eqnarray}
& \sim & (a) R\delta^{3}(k_{1}+k_{2}+k_{3}) \prod_{i=1}^{3}\left(\Delta_{a_{i}\
    -a_{i}}^{(T=0)}(t_{i}-t)\right)\nonumber\\
&  & \times \Delta_{a_{1}a_{2}}^{(T=0)}(t_{1}-t_{2})\Delta_{a_{2}a_{3}}^{(T=0)}
(t_{2}-t_{3}), \label{isolated}
\end{eqnarray}
where $R=R(a_{1},a_{3})$ denotes the contribution from the rest of the
graph and we
have used the relations in (\ref{identities}) in simplifying the
integrand. In this case we
see that the integrand depends linearly on $(a)$ and, consequently,
if we were to sum the contributions of two graphs corresponding to the
two values of this index, the sum would vanish. However, in this
theory even the individual graphs for any fixed value of the thermal
index lead to a vanishing contribution which can be seen as
follows. We note that $t$ represents an internal time coordinate which
needs to be integrated over. Using the form of the propagators in
(\ref{aminusa}) and integrating over $t$, the contribution of the
graph for any value of the ``thermal''
indices becomes proportional to
\begin{equation}
\sim
\delta^{3}(k_{1}+k_{2}+k_{3})\delta(a_{1}E_{1}+a_{2}E_{2}+a_{3}E_{3})
= 0.\label{conservation}
\end{equation}
The vanishing of this graph follows from the fact that there cannot be
a decay involving three on-shell massive particles. (Remember that $E_{i} =
\sqrt{k_{i}^{2} + m^{2}}$.) Through the application of the thermal
operator, it follows then that such a graph with an isolated circled
internal vertex will vanish at finite temperature for any value of the
thermal index.

\begin{figure}[ht!]
\begin{center}
\includegraphics[scale=0.4]{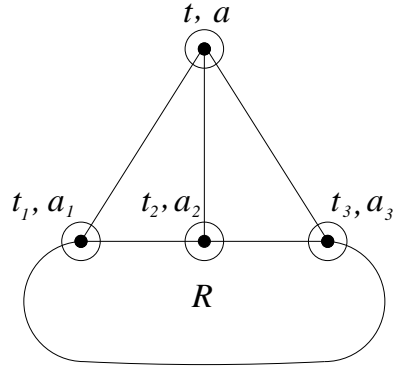}
\end{center}
\caption{A typical diagram with an isolated island of internal circled
  vertices.}\label{fe3}
\end{figure}

\begin{figure}[ht!]
\begin{center}
\includegraphics[scale=0.4]{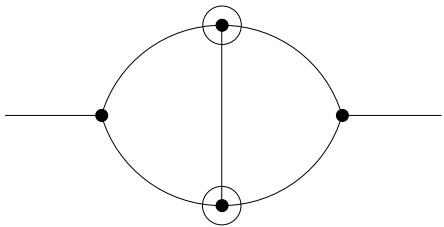}
\end{center}
\caption{A two loop self-energy graph with two internal circled
  vertices.}\label{fe4}
\end{figure}

We can now show using
the relation (\ref{basic}) that if we have a diagram with an island of internal
circled vertices, then its contribution identically vanishes
(without the use of any energy conservation) when summed over the
``thermal'' indices of all the circled vertices. This can be seen from
the graph shown in figure \eqref{fe3} as
follows. Let us assume that the vertex $(t,a)$ has the smallest time
among the internal circled vertices. Then, using (\ref{basic}) we see
that the integrand of the graph 

\noindent would have the form (\ref{isolated}) (with the last two propagators doubly circled) 
and will be linear in
the factor $(a)$. As a result,  when summed over the ``thermal''index
$a$, the integrand vanishes (as in the case of a single isolated
circled internal vertex). We note that the time coordinates of
internal vertices need to be integrated over. Therefore, there will be
time configurations for which the vertex $(t_{1},a_{1})$, for example, 
will have the shortest time among the circled vertices. In this case, the
above argument can be repeated and it will follow that the diagram
vanishes when summed over the index $a_{1}$. In this way, it is clear that
the contribution of the diagram will totally vanish when we sum over
the ``thermal'' indices of all the internal circled vertices. Once
again, since these contributions vanish identically without the use of
any relation of energy conservation, through the thermal operator, we
see that such diagrams must also vanish at finite temperature. An
example of such a diagram that can be explicitly checked to vanish
when summed over the indices of the circled vertices is the following
two loop self-energy diagram shown in figure \eqref{fe4}.

\begin{figure}[ht!]
\begin{center}
\includegraphics[scale=0.4]{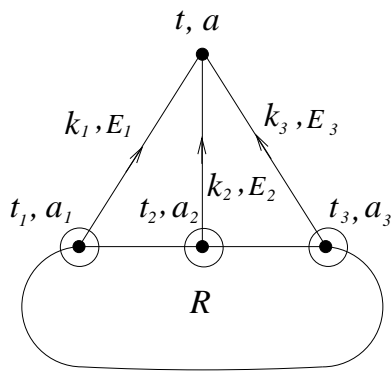}
\end{center}
\caption{A typical diagram with an isolated internal uncircled
  vertex.}\label{fe5}
\end{figure}

Let us next consider a generic diagram in $\phi^{3}$ theory shown in figure \eqref{fe5},
where there is an isolated internal
vertex that is uncircled and is connected only to circled vertices
that are internal. 

\noindent with the integrand given by (we suppress the energy
dependence in the propagator for simplicity of notation) 
\begin{eqnarray}
&\sim & \delta^{3}(k_{1}+k_{2}+k_{3})\sum_{a=\pm} (a)
\prod_{i=1}^{3}\left(\Delta_{a\ -a} (t-t_{i})\right) \nonumber\\
& \times&\left[\sum_{a_{i}} R
  (a_{1})(a_{2})(a_{3})\Delta_{a_{1}a_{2}}(\underline{t_{1}}-\underline{t_{2}}) 
\Delta_{a_{2}a_{3}}
(\underline{t_{2}}-\underline{t_{3}})\right].\nonumber\\
& & \label{uncircled}
\end{eqnarray}
The coordinate $t$ is internal and needs to be integrated over. As in
the case of an isolated circled internal vertex, if we integrate over
$t$, the factor inside the product of propagators leads to
\begin{equation}
\sim \delta^{3} (k_{1}+k_{2}+k_{3}) \delta (E_{1}+E_{2}+E_{3}) = 0,
\end{equation}
much like in (\ref{conservation}).
This shows that an isolated
internal uncircled vertex connected to internal circled vertices
vanishes for any value of the internal ``thermal''index of the
uncircled vertex. Since this graph vanishes identically at zero
temperature, through the use of the thermal operator, it follows that
such a graph will also vanish at finite temperature.

\begin{figure}[ht!]
\begin{center}
\includegraphics[scale=0.4]{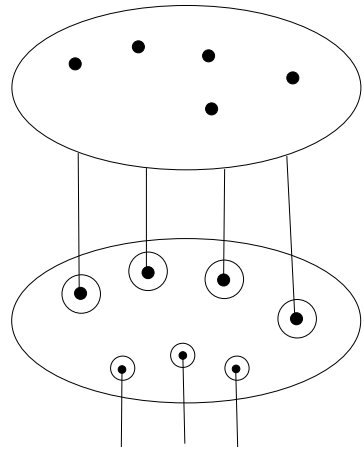}
\end{center}
\caption{A typical diagram of an isolated island of uncircled internal
  vertices connected to internal circled vertices.}\label{fe13}
\end{figure}

From the discussion above, it is clear that if in a diagram, we have
an internal uncircled vertex connected to three circled vertices, then
the diagram vanishes by energy conservation, both at zero and at
finite temperature, when we integrate over the time coordinate of the
internal uncircled vertex. Furthermore, this happens for any
distribution of the ``thermal'' indices. Let us next consider the
diagram shown in figure \eqref{fe13}, where an isolated island of uncircled
internal vertices is connected to internal circled vertices. A general
proof for the vanishing of such a graph is rather involved at finite
temperature and we refer the reader to \cite{das:book97} for
details. Here  we
summarize in a simple manner what goes into such a proof through the
application of the thermal operator.   

\begin{figure}[ht!]
\begin{center}
\includegraphics[scale=0.4]{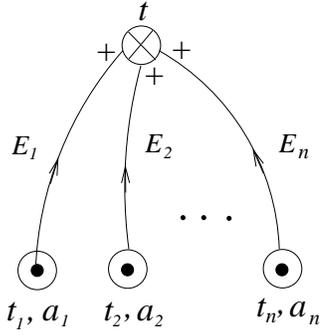}
\end{center}
\caption{An $n$-point vertex correction arising from uncircled
  vertices connected to internal circled vertices.}\label{fe12}
\end{figure}

If the
island of uncircled vertices contains at least one vertex that is
connected only to the circled vertices, then such a diagram will
again vanish, both at zero as well as at finite temperature, because
of arguments  of energy conservation given above. However, if the
island of uncircled internal vertices does not contain any vertex that
is not connected only to circled vertices, the vanishing of such a
graph at finite temperature (under the action of thermal operator)
does not follow from arguments of energy conservation as given above
for any distribution of ``thermal'' indices. For example, let us
consider for  simplicity
the case where all the vertices in the island of uncircled vertices
are of ``$+$'' type and 
are connected among 
themselves (as well as to the internal circled vertices). Integrating
over all except one of the internal time coordinate, the island of
internal uncircled vertices can be thought of as a $n$-point vertex
correction connected to internal circled vertices as shown in figure
\eqref{fe12}.

\noindent In this case, integrating over the internal time coordinate
$t$, leads to 
\begin{equation}
\sim \delta (E_{1}+E_{2}+\cdots + E_{n}),
\end{equation}
for any given distribution of ``thermal'' indices. This vanishes at
zero temperature. However, under the action of the thermal operator,
some of the $E_{i}$'s inside the delta function will change sign and,
therefore, the vanishing does not hold for individual
graphs. On the other hand, if we sum over the complete set of thermal
indices of the internal uncircled vertices, 
then the contribution is annihilated by the thermal operator
as we will show in detail in the following example.

\begin{figure}[t!]
\begin{center}
\includegraphics[scale=0.4]{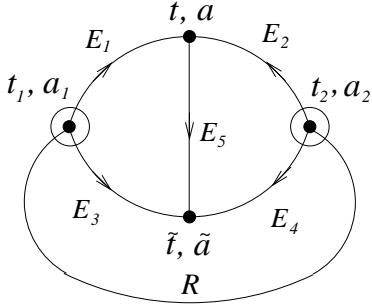}
\end{center}
\caption{A two loop self-energy insertion in a generic diagram.}\label{fe11}
\end{figure}

Let us consider a typical diagram in $\phi^{3}$ theory with an insertion of
the  two loop self-energy correction graph as shown in figure
\eqref{fe11}.  In this case, we note that 
there are two isolated uncircled internal vertices connected among themselves as well as 
to internal circled vertices (there is no internal uncircled vertex that is connected only 
to circled vertices). In this case, the contribution of the diagram can be written as
\begin{equation}
\sim \sum_{a_{1},a_{2}} (a_{1})(a_{2}) R(a_{1},a_{2}) \sum_{a,\tilde{a}} \Gamma_{a\tilde{a}}^{(T=0)},
\end{equation}
where we have identified
\begin{eqnarray}
& & \Gamma_{a\tilde{a}}^{(T=0)} = (a)(\tilde{a}) \int dt d\tilde{t}\ 
 \Delta_{a
  \tilde{a}}^{(T=0)} (t-\tilde{t}, E_{5})\nonumber\\
 & &  \times \prod_{\alpha=1}^{2} \Delta_{-a
  a}^{(T=0)}(t_{\alpha}-t,E_{\alpha})\Delta_{-\tilde{a}
  \tilde{a}}^{(T=0)} (t_{\alpha}-\tilde{t}, E_{\alpha+2}).\label{general} 
\end{eqnarray}
Using the relation (\ref{aminusa}), the integration over $t,\tilde{t}$
can be done and for various combinations of the ``thermal''  indices, the results are
\begin{widetext}
\begin{eqnarray}
\Gamma_{++}^{(T=0)} & = & -(2\pi i) \left(\prod_{i=1}^{5} \frac{1}{2E_{i}}\right) \delta(E_{1}+E_{2}+E_{3}+E_{4})\nonumber\\
& & \ \times e^{-i(E_{1}+E_{3})(t_{1}-t_{2})}\left[\frac{1}{E_{3}+E_{4}+E_{5}-i\epsilon} - \frac{1}{E_{3}+E_{4}-E_{5}+i\epsilon}\right],\nonumber\\
\Gamma_{+-}^{(T=0)} & = & - (2\pi)^{2} \left(\prod_{i=1}^{5} \frac{1}{2E_{i}}\right) \delta(E_{1}+E_{2}-E_{3}-E_{4})\nonumber\\
& & \ \times e^{-i(E_{1}-E_{3})(t_{1}-t_{2})}\ \delta(E_{3}+E_{4}+E_{5}),\nonumber\\
\Gamma_{-+}^{(T=0)} & = & - (2\pi)^{2} \left(\prod_{i=1}^{5} \frac{1}{2E_{i}}\right) \delta(E_{1}+E_{2}-E_{3}-E_{4})\nonumber\\
& & \ \times e^{i(E_{1}-E_{3})(t_{1}-t_{2})}\ \delta(E_{3}+E_{4}+E_{5}),\nonumber\\
\Gamma_{--}^{(T=0)} & = & (2\pi i) \left(\prod_{i=1}^{5} \frac{1}{2E_{i}}\right) \delta(E_{1}+E_{2}+E_{3}+E_{4})\nonumber\\
& & \ \times e^{i(E_{1}+E_{3})(t_{1}-t_{2})}\left[\frac{1}{E_{3}+E_{4}+E_{5}+i\epsilon} - \frac{1}{E_{3}+E_{4}-E_{5}-i\epsilon}\right].\label{comp}
\end{eqnarray}
\end{widetext}
It is clear from the structures in (\ref{comp}) that every single
component vanishes at zero temperature as a consequence of an energy
conserving delta function. We can now apply the thermal operator to
these components and explicitly verify that at finite temperature the
components no longer vanish individually. On the other hand, if we sum
over the ``thermal''  indices of the internal uncircled vertices
(namely, sum over all the components in (\ref{comp}) after applying
the thermal operator), then the sum identically vanishes without the
use of energy conservation. We have  checked this directly, which is
tedious, but there is a simpler and more elegant
way of seeing this cancellation as follows. 

Let us recall (see (\ref{factorization})) that applying the thermal operator simply changes a zero temperature propagator to a finite temperature one. Thus, applying the thermal operator to (\ref{general}) we can write 
\begin{widetext}
\begin{eqnarray}
\Gamma_{++}^{(T)} & = & \int dt d\tilde{t} \prod_{\alpha=1}^{2} \Delta_{-+}^{(T)} (t_{\alpha}-t,E_{\alpha}) \Delta_{-+}^{(T)} (t_{\alpha}-\tilde{t},E_{\alpha+2}) \Delta_{++}^{(T)} (t-\tilde{t},E_{5}),\nonumber\\
\Gamma_{+-}^{(T)} & = & -\int dt d\tilde{t} \prod_{\alpha=1}^{2} \Delta_{-+}^{(T)} (t_{\alpha}-t,E_{\alpha}) \Delta_{+-}^{(T)} (t_{\alpha}-\tilde{t},E_{\alpha+2}) \Delta_{+-}^{(T)} (t-\tilde{t},E_{5}),\nonumber\\
\Gamma_{-+}^{(T)} & = & -\int dt d\tilde{t} \prod_{\alpha=1}^{2} \Delta_{+-}^{(T)} (t_{\alpha}-t,E_{\alpha}) \Delta_{-+}^{(T)} (t_{\alpha}-\tilde{t},E_{\alpha+2}) \Delta_{-+}^{(T)} (t-\tilde{t},E_{5}),\nonumber\\
\Gamma_{--}^{(T)} & = & \int dt d\tilde{t} \prod_{\alpha=1}^{2} \Delta_{+-}^{(T)} (t_{\alpha}-t,E_{\alpha}) \Delta_{+-}^{(T)} (t_{\alpha}-\tilde{t},E_{\alpha+2}) \Delta_{--}^{(T)} (t-\tilde{t},E_{5}).
\end{eqnarray}
\end{widetext}
We can now use the KMS condition (\ref{KMS})
\begin{eqnarray}
\Delta_{-+}^{(T)} (t, E) & = &  \Delta_{+-}^{(T)}
(t+i\beta,E),\nonumber\\
\Delta_{+-}^{(T)} (t, E) & = & \Delta_{-+}^{(T)} (t-i\beta, E),
\end{eqnarray}
in the first three terms to change all the prefactors to be products
of $\Delta_{+-}^{(T)}$ (we also change the last factor in the middle
two terms using these relations). Furthermore, since $t,\tilde{t}$ are internal
coordinates that are being integrated over, in the first three terms we
can make the shifts
\begin{equation}
t\rightarrow t+i\beta,\quad \tilde{t}\rightarrow \tilde{t}+i\beta,
\end{equation}
as is necessary
to write the sum over the internal thermal indices of the uncircled vertices as
\begin{eqnarray}
\sum_{a,\tilde{a}} \Gamma_{a\tilde{a}}^{(T)} & = & \int dt d\tilde{t} \prod_{\alpha=1}^{2} \Delta_{+-}^{(T)} (t_{\alpha}-t,E_{\alpha}) \Delta_{+-}^{(T)} (t_{\alpha}-\tilde{t},E_{\alpha+2})\nonumber\\
& & \quad \times \left(\sum_{a,\tilde{a}} (a)(\tilde{a}) \Delta_{a \tilde{a}}^{(T)} (t-\tilde{t},E_{5})\right).
\end{eqnarray}
The last sum is easily seen to vanish using (\ref{identity1}). Thus, we see that even though individual diagrams vanish at zero temperature because of energy conservation, at finite temperature the vanishing holds only when we sum over the thermal indices of the internal uncircled vertices.

\begin{figure}[ht!]
\begin{center}
\includegraphics[scale=0.4]{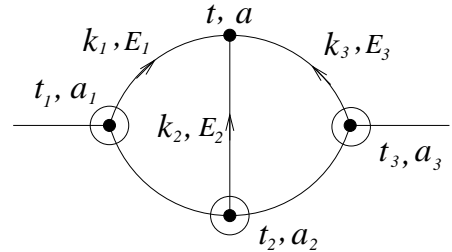}
\end{center}
\caption{A two loop self-energy diagram with an internal uncircled
  vertex connected to external circled vertices.}\label{fe7}
\end{figure}

In deriving these relations, we have assumed that the internal uncircled
vertex or the island of uncircled vertices are connected to internal
circled vertices. If any of the circled vertices is external, then our
argument does not go through and, in fact, such diagrams would not in
general vanish since they are needed to give a cutting description to
the diagrams. However, some such diagrams where the internal uncircled
vertex is connected to external circled vertex may identically vanish
from conservation laws. For example, let us consider the following two
loop self-energy diagram in the $\phi^{3}$ theory, shown in figure
\eqref{fe7}.

\noindent In this case, we note that for any value of the thermal
index $a$, the integrand has the form
\begin{eqnarray}
&\sim& R \delta^{3} (k_{1}+k_{2}+k_{3}) \prod_{i} \Delta_{a\ -a}
(t-t_{i}, E_{i})\nonumber\\
& = & R \delta^{3} (k_{1}+k_{2}+k_{3})
\prod_{i}\frac{e^{i(a)E_{i}(t-t_{i})}}{2E_{i}}
\end{eqnarray}
When integrated over the internal time coordinate, the result has the
form
\begin{equation}
\sim \delta^{3} (k_{1}+k_{2}+k_{3}) \delta (E_{1}+E_{2}+E_{3}) = 0.
\end{equation}
As we have noted earlier, the vanishing of this graph follows from the
fact that there cannot be
a decay involving three on-shell massive particles. This conclusion
holds even if one of the
$E_{i}$'s changes sign in the delta function and so the vanishing of
these individual graphs continues to hold at finite temperature (as
can be easily seen from the action of the thermal operator).

Thus, we see that in this doubled theory at zero temperature, when the
internal ``thermal'' indices are summed, the nontrivial graphs
contributing to the imaginary part in the momentum space consist of
diagrams where there are regions of circled and uncircled vertices
connected to the external vertices in a continuous manner such that a
cutting description holds. Furthermore, through the application of the
thermal operator (since every propagator factors into the same thermal
operator that is independent of time), it follows that a cutting
description of the graphs holds even at finite temperature in a
completely parallel manner. The cutting rules for this theory at
finite temperature and $\mu=0$ are
already well known. However, this gives a simpler derivation of the
cutting description through the application of the thermal operator.

\subsection{Example}

As an example of the cutting description through the thermal operator,
let us calculate the imaginary part of the one loop retarded self-energy at
finite temperature for the $\phi^{3}$ theory. This imaginary part has
been calculated from various points of view. Here we merely give a
brief derivation of this to illustrate the application of the thermal
operator representation.

In the case of the
self-energy, Eq. (\ref{imaginary}) graphically takes the form
shown in figure \eqref{fx1}.

\begin{widetext}
\begin{center}
\begin{figure}[ht!]
\includegraphics[scale=0.2]{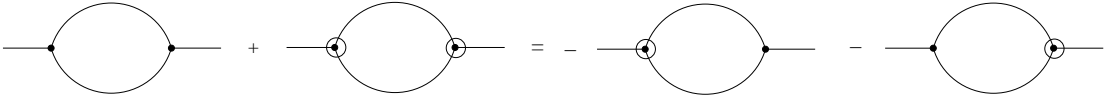}
\caption{Graphical representation of Eq. (17) for a one loop
  self-energy diagram.}\label{fx1}
\end{figure}
\end{center}
\end{widetext}

\begin{figure}[ht!]
\begin{center}
\includegraphics[scale=0.4]{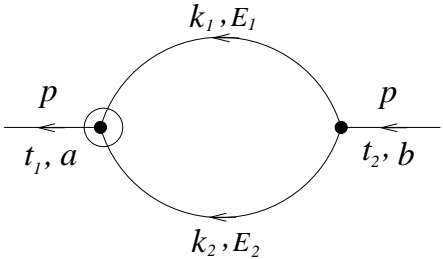}
\end{center}
\caption{The one loop self-energy graph leading to the imaginary
  part.}\label{fx2}
\end{figure}

\noindent Furthermore, if we are interested only in the retarded self-energy,
the second diagram with the circled vertex on the right can be shown
to add up to zero. Consequently, the nontrivial diagram that would
contribute to the imaginary part of the retarded self-energy at one loop has
the form shown in figure \eqref{fx2}.

\noindent Ignoring the momentum integration, (namely, $\int
\frac{d^{3}k_{1} d^{3}k_{2}}{(2\pi)^{6}}\ (2\pi)^{3} \delta^{3}
(k_{1}+k_{2}-p)$), the diagram leads to
\begin{eqnarray}
-i\overline{\Pi}_{ab}^{(T=0)} & = & \frac{(-ig)(ig)(a)(b)}{2}
\prod_{i=1}^{2}\Delta_{-b\ b}
(t_{1}-t_{2},E_{i})\nonumber\\
 & = & \frac{(a)(b) g^{2}}{2} \frac{1}{4E_{1}E_{2}}\ e^{-i(b)
   (E_{1}+E_{2})(t_{1}-t_{2})},\label{piab}
\end{eqnarray}
where we have used the compact representation of the propagators in
(\ref{aminusa}) as well as the fact that the two point function is
identified with 
$-i\overline{\Pi}$ where $\overline{\Pi}$ denotes the appropriate self-energy. 
Since the retarded self-energy at zero temperature is
given by 
\begin{equation}
\Pi_{\rm R}^{(T=0)} = \Pi_{++}^{(T=0)} + \Pi_{+-}^{(T=0)},\label{retarded}
\end{equation}
using (\ref{piab}) and taking the Fourier transform with respect to
the time coordinate, we obtain the imaginary part of the retarded
self-energy in momentum space to be
\begin{eqnarray}
\lefteqn{{\rm Im} \Pi_{\rm R}^{(T=0)} (p)= - \frac{1}{2}
  (-i\overline{\Pi}_{\rm R}^{(T=0)} (p) = \frac{i}{2}
  \overline{\Pi}_{\rm R}^{(T=0)} (p)}\nonumber\\
& = & - \frac{1}{2}\int dt
\frac{g^{2}}{8E_{1}E_{2}}\left(e^{i(p_{0}-E_{1}-E_{2})t} -
  e^{i(p_{0}+E_{1}+E_{2})t}\right)\nonumber\\
 & = &\!\! -\frac{(2\pi)g^{2}}{16E_{1}E_{2}}\left(\delta(p_{0}-E_{1}-E_{2})
   - \delta (p_{0}+E_{1}+E_{2})\right).\label{imaginaryretarded0}
\end{eqnarray}
Applying the relevant thermal operator for the graph \cite{silvana1,Brandt:2006rv}
\begin{equation}
{\cal O}^{(T)} = \prod_{i=1}^{2} \left(1 + n(E_{i}) (1-
  S(E_{i}))\right),
\end{equation}
we obtain the imaginary part of the retarded self energy at one loop
at finite temperature to be 
\begin{eqnarray}
\lefteqn{{\rm Im} \Pi_{\rm R}^{(T)} (p)}\nonumber\\
& = & -
\frac{(2\pi)g^{2}}{16E_{1}E_{2}}\Big[(1+ n(E_{1})+n(E_{2}))\nonumber\\
& & \quad \times(\delta(p_{0}-E_{1}-E_{2})
- \delta (p_{0} +E_{1} +E_{2}))\nonumber\\
 & & \quad + (n(E_{1})-n(E_{2}))\nonumber\\
& & \quad \times (\delta (p_{0}+E_{1}-E_{2})-\delta
 (p_{0}-E_{1}+E_{2}))\big],\label{imaginaryretarded}
\end{eqnarray}
which is well known in the literature.

\section{Cutting Rules at Finite Temperature and $\mu\neq 0$}

The cutting rules at finite temperature in the absence of a chemical
potential are well known in the closed time path formalism. In the
earlier section, we have given a simple derivation of these rules
through the application of the thermal operator on the cutting
description of a zero temperature theory with doubled degrees of
freedom. In this section we will derive the cutting rules for a theory
at finite temperature with a nonzero chemical potential through the
application of the thermal operator (which we have also verified
directly) and to the best of our knowledge, this has not been done in
the closed time path formalism. As we will see, the proof of a cutting
description in this case will be quite parallel to that discussed in
the last section. Therefore, instead of repeating arguments, we will
give only the essential details in this section.

Let us consider for simplicity a toy model of an interacting theory of a real 
and  a complex scalar field described by the Lagrangian density
\begin{eqnarray}
{\cal L} & = & \frac{1}{2}\partial_{\mu}\sigma \partial^{\mu}\sigma -
\frac{m^{2}}{2} \sigma^{2} +
\left((\partial_{t}-i\mu)\phi\right)^{*}(\partial_{t}-i\mu)\phi\nonumber\\
& & \quad - (\vec{\nabla}\phi)^{*}\cdot \vec{\nabla}\phi -
M^{2}\phi^{*}\phi - g \sigma\phi^{*}\phi,
\end{eqnarray}
where $\mu$ stands for the chemical potential of the complex scalar
field and is assumed to have a value $\mu <M$. For the real scalar field, there is no chemical potential and
the components of the thermal propagator in the closed time path
formalism factorize through a thermal operator as discussed in
Eqs. (\ref{factorization}, \ref{tor}). For the complex scalar field
with a chemical potential, however, the components of the propagator
in the closed time path formalism are more complicated. In the momentum space, they can be written as
\begin{eqnarray}
G_{++}^{(T,\mu)} (p) & = & \frac{i}{(p_{0}+\mu)^{2} -
  E^{2}+i\epsilon}\nonumber\\
& & \quad + 2\pi n^{(\mu)}(p_{0}) \delta ((p_{0}+\mu)^{2}-E^{2}),\nonumber\\
G_{+-}^{(T,\mu)} (p) & = & 2\pi\left(\theta(-p_{0}-\mu) +
  n^{(\mu)}(p_{0})\right)\nonumber\\
& & \quad\times \delta ((p_{0}+\mu)^{2}-E^{2}),\nonumber\\
G_{-+}^{(T,\mu)} (p) & = & 2\pi\left(\theta(p_{0}+\mu) + n^{(\mu)}
  (p_{0})\right)\nonumber\\
& & \quad\times\delta ((p_{o}+\mu)^{2}-E^{2}),\nonumber\\
G_{--}^{(T,\mu)} (p) & = & -
\frac{i}{(p_{0}+\mu)^{2}-E^{2}-i\epsilon}\nonumber\\
& & \quad + 2\pi n^{(\mu)}(p_{0}) \delta ((p_{0}+\mu)^{2}-E^{2}),\label{mupprop}
\end{eqnarray}
where
\begin{equation}
E = \sqrt{\vec{p}^{\ 2} + M^{2}},\quad n^{(\mu)}(p_{0}) = n (p_{0}{\rm sgn} (p_{0}+\mu)).
\end{equation}
In the mixed space, they take the forms \cite{silvana1}
\begin{eqnarray}
G_{++}^{(T,\mu)}(t,E) & = &
\frac{1}{2E}\left((\theta(t)+n_{-}(E))e^{-iE_{-}t}\right.\nonumber\\
& & \qquad \left. +
  (\theta(-t)+n_{+}(E))e^{iE_{+}t}\right),\nonumber\\
G_{+-}^{(T,\mu)}(t,E) & = & \frac{1}{2E}\left(n_{-}(E)
  e^{-iE_{-}t} + (1 + n_{+}(E)) e^{iE_{+}t}\right),\nonumber\\
G_{-+}^{(T,\mu)}(t,E) & = & \frac{1}{2E}\left((1+n_{-})
  e^{-iE_{-}t} + n_{+}(E) e^{iE_{+}t}\right),\nonumber\\
G_{--}^{(T,\mu)}(t,E) & = &
\frac{1}{2E}\left((\theta(t)+n_{+}(E))e^{iE_{+}t}\right.\nonumber\\
& & \qquad \left. +
  (\theta(-t)+n_{-})e^{-iE_{-}t}\right),\label{muprop}
\end{eqnarray}
where we have defined
\begin{equation} 
E_{\pm} = E\pm \mu,\quad n_{\mp}(E) = n (E\mp \mu) =
\frac{1}{e^{\beta(E\mp \mu)} - 1}. 
\end{equation}
In the zero temperature limit, the components of the propagator take
the form
\begin{eqnarray}
G_{++}^{(T=0,\mu)}(t,E) & = & \frac{1}{2E}\left(\theta(t)e^{-iE_{-}t}
  + \theta(-t)e^{iE_{+}t}\right),\nonumber\\
G_{+-}^{(T=0,\mu)}(t,E) & = & \frac{1}{2E}\ e^{iE_{+}t},\nonumber\\
G_{-+}^{(T=0,\mu)}(t,E) & = & \frac{1}{2E}\ e^{-iE_{-}t},\nonumber\\
G_{--}^{(T=0,\mu)}(t,E) & = & \frac{1}{2E}\left(\theta(t) e^{iE_{+}t}
  + \theta(-t) e^{-iE_{-}t}\right).\label{zeroTmuprop}
\end{eqnarray}

As is clear from (\ref{muprop}), in the presence of a chemical
potential, the components of the thermal propagator are more
complicated mainly because the distribution function for the positive
and the negative frequency terms in the propagator are different. In
this case, a simple factorization as in
(\ref{factorization},\ref{tor}) in terms of the simple reflection
operator $S(E)$ alone does not work. Rather, a time independent
factorization of the propagator and, therefore, a thermal
representation for any finite temperature graph can be obtained if we
introduce an additional operator $\hat{N}^{(T,\mu)}(E)$ such that
\begin{equation}
\hat{N}^{(T,\mu)}(E) f(E_{\mp}) = n_{\mp}(E) f(E_{\mp}).\label{relation}
\end{equation}
In this case, we can write the components of the thermal propagator in
a factorized manner as
\begin{equation}
G_{ab}^{(T,\mu)}(t,E) = {\cal O}^{(T,\mu)}(E)
G_{ab}^{(T=0,\mu)}(E),\label{mufactorization}
\end{equation}
where
\begin{equation}
{\cal O}^{(T,\mu)}(E) = 1 + \hat{N}^{(T,\mu)}(E) (1 -
S(E)).\label{mutor}
\end{equation}
The action of this additional operator has already been 
discussed in \cite{Inui:2006jf,Brandt:2006vy}
to which we refer the readers for more details.

Given the mixed space propagators of the zero temperature theory with
doubled degrees of freedom, one look at their positive and
negative frequency decomposition as in (\ref{frequencydecomp}). This
leads to the set of circled propagators as in (\ref{circledprop}) both
at zero as well as finite temperatures. (We give the spectral
representation for the components of the propagator at finite
temperature in  appendix {\bf B}.) It is easy to see from the definition of
the circled propagators that the set of circled propagators at finite
temperature factorize into that at zero temperature and the thermal
operator (\ref{mutor}), namely,
\begin{eqnarray}
G_{ab}^{(T,\mu)}(t_{1}-t_{2},E) & = & {\cal O}^{(T,\mu)}(E)
G_{ab}^{(T=0,\mu)}(t_{1}-t_{2},E),\nonumber\\
G_{ab}^{(T,\mu)}(t_{1}-\underline{t_{2}},E) & = & {\cal
  O}^{(T,\mu)}(E)
G_{ab}^{(T=0,\mu)}(t_{1}-\underline{t_{2}},E),\nonumber\\
G_{ab}^{(T,\mu)}(\underline{t_{1}}-t_{2},E) & = & {\cal
  O}^{(T,\mu)}(E)
G_{ab}^{(T=0,\mu)}(\underline{t_{1}}-t_{2},E),\nonumber\\
G_{ab}^{(T,\mu)}(\underline{t_{1}}-\underline{t_{2}},E) & = & {\cal
  O}^{(T,\mu)}(E)
G_{ab}^{(T=0,\mu)}(\underline{t_{1}}-\underline{t_{2}},E).
\nonumber \\
\end{eqnarray}
We note here that the propagator where both ends are circled
corresponds to the anti time ordered propagator which can be easily
seen to be the complex conjugate of the original propagator in the
momentum space. (It is interesting to point out here that in the
presence of the chemical potential, the doubly circled propagator is
not the complex conjugate even for the ``$\pm\pm$'' components in the
mixed space unlike the case when $\mu=0$.) We can also introduce the
circled vertices as in (\ref{circledvertices}) in this theory.

From the definition of the set of circled propagators, identities such
as (\ref{lte},\ref{lte1}) can be easily seen to hold in the presence
of a chemical potential and, therefore, it follows that the largest
time equation (see (\ref{lte2})) also holds in this case. Through the
application of the thermal operator, the largest time equation also
holds at finite temperature in the presence of a chemical
potential. Namely, if we add to a graph with the largest time
coordinate uncircled, a corresponding graph with the largest time
coordinate circled, the sum identically vanishes. Incidentally, since
identities such as (\ref{lte},\ref{lte1}) can be checked directly to
hold for the components of the propagator at finite temperature with
$\mu\neq 0$, the largest time equation can also be checked to hold
directly (independent of the proof through the application of the
thermal operator). From the largest time equation, it then follows
that the diagrams still satisfy the identity (\ref{imaginary}) so that
the imaginary part of a diagram in momentum space can be given a
diagrammatic representation. 

To obtain a cutting description, we proceed as in the last
section. First, it can be checked as in the case of $\mu=0$ that there
are only two basic independent components of the propagator, namely,
$G_{+-}^{(T=0,\mu)}, G_{-+}^{(T=0,\mu)}$ such that we can express all
the components of the propagators including the circled ones as
($a,b=\pm$) 
\begin{eqnarray}
G_{ab}^{(T=0,\mu)} (t_{1}-\underline{t_{2}},E) & = & G_{a\ -a}^{(T=0,\mu)}(t_{1}-t_{2},E),\nonumber\\
G_{ab}^{(T=0,\mu)}(\underline{t_{1}}-t_{2},E) & = & G_{-b\ b}^{(T=0,\mu)}(t_{1}-t_{2},E),\nonumber\\
G_{ab}^{(T=0,\mu)}(t_{1}-t_{2},E) & = & \theta (t_{1}-t_{2}) G_{-b\ b}^{(T=0,\mu)} (t_{1}-t_{2},E)\nonumber\\
 &+& \theta(t_{2}-t_{1}) G_{a\ -a}^{(T=0,\mu)}(t_{1}-t_{2},E),\nonumber\\
 G_{ab}^{(T=0,\mu)}(\underline{t_{1}}-\underline{t_{2}},E) & = & \theta(t_{1}-t_{2}) G_{a\ -a}^{(T=0,\mu)}(t_{1}-t_{2},E)\nonumber\\
  &+& \theta(t_{2}-t_{1}) G_{-b\
    b}^{(T=0,\mu)}(t_{1}-t_{2},E).\nonumber\\
& & \label{simpleid}
  \end{eqnarray}
 From this relation it follows that much like (\ref{identity1}), for
 any circling of the time coordinates $t_{1},t_{2}$, we have 
 \begin{equation}
 \sum_{a,b=\pm} (a)(b) G_{ab}^{(T=0,\mu)} (t_{1}-t_{2},E) = 0.
 \end{equation}
Let us also note for completeness that the basic components of the
propagator $G_{\pm\mp}^{(T=0,\mu)}$ can be written in a compact form
(see (\ref{aminusa})) as
\begin{equation}
G_{a\ -a}^{(T=0,\mu)} (t_{1}-t_{2},E) = \frac{1}{2E} e^{i (a)
  E_{(a)}(t_{1}-t_{2})}.\label{muaminusa}
\end{equation}

Since all the basic identities one needs to prove a cutting
description of the imaginary part of a diagram (in the momentum space)
in the case $\mu=0$ also hold for $\mu\neq 0$, we can go through the
discussions of the last section. However, without repeating the
arguments, we simply conclude that in the presence of a chemical
potential, a cutting description for the imaginary part of a graph
holds in the  zero temperature theory with doubled degrees of freedom,
when summed over the ``thermal" indices of the internal vertices in a
graph. Through the application of the thermal operator, we then
conclude that such a description also holds at finite temperature with
$\mu\neq 0$. We note here that independent of the thermal operator
argument, one can directly verify that all the relevant identities
hold for the components of the propagator at finite temperature and
$\mu\neq 0$ so that one can also prove a cutting description for any
graph at finite temperature directly (which we have done). However,
the power of the thermal operator representation is that once the
cutting description is shown to hold at zero temperature in the theory
with a doubled degrees of freedom, it automatically holds at finite
temperature. 
 
\subsection{Example}

\begin{figure}[ht!]
\begin{center}
\includegraphics[scale=0.4]{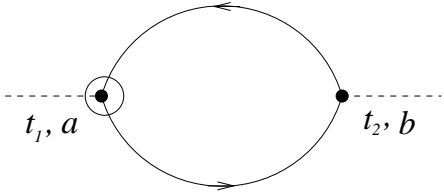}
\end{center}
\caption{One loop diagram leading to the imaginary part of the self
  energy. The dashed lines represent the neutral scalar field and the
  solid lines the complex scalar field with the arrow representing the
  direction of the charge flow.}\label{fx3}
\end{figure}

As an application of the cutting rules in the presence of a chemical
potential, let us again calculate the imaginary part of the
self-energy for the real scalar particle at one loop in this
theory. Let us note that in the mixed space, the two point function
shown in figure \eqref{fx3} has a very simple form (we follow the same notation as in (\ref{piab}))

\begin{eqnarray}
\lefteqn{-i\overline{\Pi}_{ab}^{(T=0,\mu)}}\nonumber\\
 &=& \frac{g^{2}(a)(b)}{2} G_{b\
  -b}^{(T=0,\mu)}(t_{2}-t_{1},E_{1}) G_{-b\
  b}^{(T=0,\mu)}(t_{1}-t_{2},E_{2})\nonumber\\
& = & \frac{g^{2} (a)(b)}{8E_{1}E_{2}} e^{-i(b) (E_{1 (b)} + E_{2
    (-b)}) (t_{1}-t_{2})},\label{mupiab}
\end{eqnarray}
where we have used the representation given in
(\ref{muaminusa}). Since the retarded self-energy is defined as (see
(\ref{retarded}))  
\begin{equation}
\Pi_{\rm R}^{(T=0,\mu)} = \Pi_{++}^{(T=0,\mu)} +
\Pi_{+-}^{(T=0,\mu)},\label{muretarded} 
\end{equation}
we obtain from (\ref{mupiab})
\begin{eqnarray}
\lefteqn{-i\overline{\Pi}_{\rm R}^{(T=0,\mu)} (t_{1}-t_{2})}\nonumber\\
& =&
\frac{g^{2}}{8E_{1}E_{2}}\left(e^{-i(E_{1 +}+E_{2 -})(t_{1}-t_{2})} -
  e^{i(E_{1 -}+E_{2 +})(t_{1}-t_{2})}\right).\nonumber\\
& & 
\end{eqnarray}
Taking the Fourier transform of this, we obtain the imaginary part of
the retarded self-energy to be (as in the earlier example, we are
ignoring the momentum
integration and the factors associated with them, namely, $\int
\frac{d^{3}k_{1}d^{3}k_{2}}{(2\pi)^{6}} (2\pi)^{3}\delta^{3}(k_{1}+k_{2}-p)$)
\begin{eqnarray}
\lefteqn{{\rm Im} \Pi_{\rm R}^{(T=0,\mu)} (p) =  \frac{i}{2}
\overline{\Pi}_{\rm R}^{(T=0,\mu)} (p) =
-\frac{(2\pi)g^{2}}{16E_{1}E_{2}}}\nonumber\\ 
& & \times\left(\delta(p_{0}-E_{1 +}-E_{2
    -}) - \delta (p_{0}+E_{1 -} + E_{2
    +})\right).\label{imaginaryretarded1} 
\end{eqnarray}
We note that the chemical potential in a loop cancels out completely
which is also reflected in the fact that if we are to substitute the
expressions for $E_{i \pm}$ inside the delta functions, there will be
no $\mu$ dependence in the retarded self-energy. This is, in fact, the
correct result and is what we should do if we are interested in the
imaginary part of the retarded self-energy at zero
temperature. However, as explained in 
\cite{Inui:2006jf,Brandt:2006vy}, to obtain the temperature
dependent terms through the application of the thermal operator in
(\ref{mutor}) using the relation (\ref{relation}), we should use the
explicit expressions for $E_{i \pm}$ only after the application of
the thermal operator. Using the thermal operator, we obtain the
imaginary part of the retarded self-energy to be
\begin{eqnarray}\label{relation2}
\lefteqn{{\rm Im} \Pi_{\rm R}^{(T,\mu)} (p)  = {\cal O}^{(T,\mu)}(E_{1}){\cal
  O}^{(T,\mu)} (E_{2}) {\rm Im} \Pi_{\rm R}^{(T=0,\mu)} (p)}\nonumber\\
& = & -
\frac{(2\pi)g^{2}}{16E_{1}E_{2}}\big[(1+n_{+}(E_{1})+n_{-}(E_{2}))
  \delta (p_{0}-E_{1}-E_{2})\nonumber\\
& & \quad - (1+n_{-}(E_{1})+n_{+}(E_{2}))\delta
(p_{0}+E_{1}+E_{2})\nonumber\\
& & \quad -
(n_{+}(E_{1})-n_{+}(E_{2}))\delta(p_{0}-E_{1}+E_{2})\nonumber\\
& & \quad + (n_{-}(E_{1})-n_{-}(E_{2}))\delta
  (p_{0}+E_{1}-E_{2})\big].
\end{eqnarray} 
This is easily seen to reduce to (\ref{imaginaryretarded}) when
$\mu=0$ and has all the symmetry properties of a retarded self-energy.

We would like to point out here that with the use of the circled
propagators and vertices, one can not only calculate the imaginary
part of the retarded self-energy, but also the complete retarded
self-energy as well as the appropriate dispersion relation at any
temperature (including zero temperature). Let us note using the
largest time equation (analogous to (\ref{lte1}) for the self-energy)
that
\begin{equation}
\theta (t_{1}-t_{2})\left(\Pi_{ab}^{(T=0,\mu)} (t_{1}-t_{2}) +
  \Pi_{ab}^{(T=0,\mu)}(\underline{t_{1}}-t_{2})\right) = 0,\label{relation1}
\end{equation}
which can also be represented graphically as
\begin{widetext}
\begin{equation}\label{xxx}
\theta(t_{1}-t_{2})\left[
\begin{array}{c}
\includegraphics[scale=0.4]{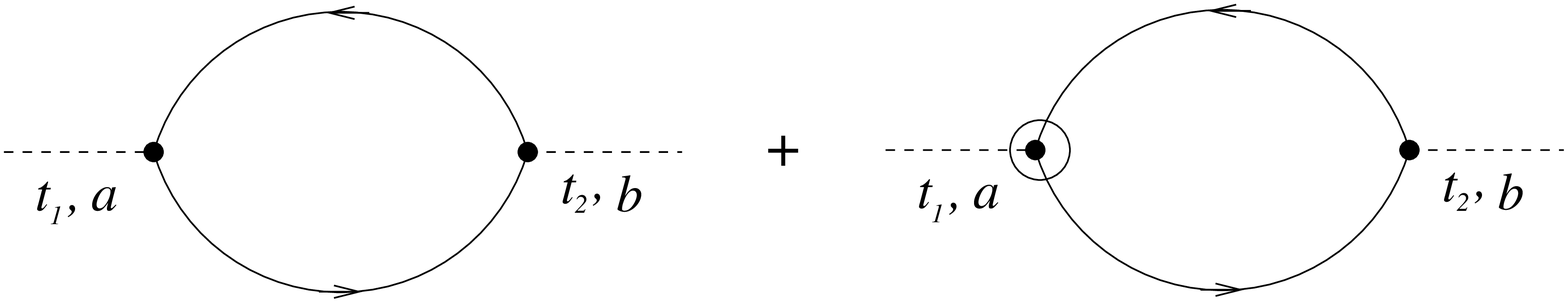}
\end{array}\right] = 0.
\end{equation}
\end{widetext}

From the definition of the retarded self-energy in (\ref{muretarded})
as well as the fact that the retarded self-energy (by definition) is
proportional to $\theta (t_{1}-t_{2})$, it follows that
\begin{eqnarray}
\lefteqn{\Pi_{\rm R}^{(T=0,\mu)} (t_{1}-t_{2}) = \theta(t_{1}-t_{2})
\Pi_{\rm R}^{(T=0,\mu)} (t_{1}-t_{2})}\nonumber\\
& = &
\theta(t_{1}-t_{2})\left(\Pi_{++}^{(T=0,\mu)}(t_{1}-t_{2})+\Pi_{+-}^{(T=0,\mu)}(t_{1}-t_{2})\right)\nonumber\\
& = &
-\theta(t_{1}-t_{2})\left(\Pi_{++}^{(T=0,\mu)}(\underline{t_{1}}-t_{2})
  + \Pi_{+-}^{(T=0,\mu)}(\underline{t_{1}}-t_{2})\right)\nonumber\\
& = & - \theta(t_{1}-t_{2}) \overline{\Pi}_{\rm R}^{(T=0,\mu)}
(t_{1}-t_{2}).\label{fullretarded}
\end{eqnarray}
Here we have used (\ref{relation1}) in the intermediate step and have
identified the appropriate graphs with
$(-i\overline{\Pi}^{(T=0,\mu)})$  defined earlier. We recall that the
Fourier transform
of $(-i\overline{\Pi})$ is related to the imaginary part of the retarded
self-energy in momentum space (see (\ref{imaginaryretarded0}) or
(\ref{imaginaryretarded1})). Therefore, (\ref{fullretarded}) gives a
method for calculating the complete retarded self-energy using the
circled propagators and vertices at zero temperature which can then be
extended to finite temperature through the application of the thermal
operator. In fact, using the integral representation for the theta
function,
\begin{equation}
\theta (t_{1}-t_{2}) = \frac{1}{2\pi i} \int dq_{0}\ \frac{e^{iq_{0}
    (t_{1}-t_{2})}}{q_{0}-i\epsilon},
\end{equation}
and taking the Fourier transform of (\ref{fullretarded}), we obtain
\begin{eqnarray}
\Pi_{\rm R}^{(T=0,\mu)} (p) &=& -\frac{1}{2\pi i} \int dq_{0}\
\frac{\overline{\Pi}_{\rm R}^{(T=0,\mu)}
  (q_{0}+p_{0},\vec{p})}{q_{0}-i\epsilon}\nonumber\\
 & = & \frac{1}{\pi} \int dq_{0}\ \frac{{\rm Im} \Pi_{\rm
     R}^{(T=0,\mu)}(q_{0}+p_{0},\vec{p})}{q_{0}-i\epsilon}.
\end{eqnarray}
This, therefore, leads to the dispersion relation for the retarded
self-energy at zero temperature and through the application of the
thermal operator, it follows that the dispersion relation holds even
at finite temperature, namely, 
\begin{equation}
\Pi_{\rm R}^{(T,\mu)}(p) = \frac{1}{\pi} \int dq_{0}\ \frac{{\rm Im}
  \Pi_{\rm R}^{(T,\mu)} (q_{0}+p_{0},\vec{p})}{q_{0}-i\epsilon}.
\end{equation}
Using the imaginary part of the retarded self-energy from
\eqref{relation2} and carrying out the delta function
 integrations, we 
obtain the complete one loop  retarded self-energy in momentum space at finite
temperature and chemical potential to be (of course, we must still perform the
momentum integrations alluded to earlier)
\begin{eqnarray}
\Pi_{\rm R}^{(T,\mu)} (p) & = & - \frac{g^{2}}{8E_{1}E_{2}}\left[
  \frac{1 + n_{+}(E_{1}) +
    n_{-}(E_{2})}{E_{1}+E_{2}-p_{0}-i\epsilon}\right.\nonumber\\
& & \quad +
\frac{1+n_{-}(E_{1})+n_{+}(E_{2})}{E_{1}+E_{2}+p_{0}+i\epsilon}\nonumber\\
& & \quad -
\frac{n_{+}(E_{1})-n_{+}(E_{2})}{E_{1}-E_{2}-p_{0}-i\epsilon}\nonumber\\
& & \quad \left. -
  \frac{n_{-}(E_{1})-n_{-}(E_{2})}{E_{1}-E_{2}+p_{0}+i\epsilon}\right].
\end{eqnarray}
This expression is exact and can be used to compute $\Pi_{\rm
  R}^{(T,\mu)} (p)$ in various limits of physical interest.

\section{Summary}

In this paper, 
we have systematically studied the interesting question
of cutting rules at finite temperature as an application of the
thermal operator representation. The thermal operator relates
in a direct manner the finite temperature graphs to those of the zero
temperature theory. Thus, we have studied first 
a zero temperature scalar theory with
doubled degrees of freedom (that can be obtained from the zero
temperature limit of the finite temperature theory in the closed time
path formalism). 
We have given an alternative algebraic derivation
of the largest time equation for this theory. We have derived the
cutting description at zero temperature and then through the action of the
thermal operator we have shown that the cutting description also holds
at finite temperature and zero/finite chemical potential. As an example, we have
calculated the imaginary part of the one loop retarded self-energy at
finite temperature and zero/finite chemical potential. We have also
shown how the circled propagators and vertices can be used to obtain
the dispersion relation as well as the full retarded self-energy of
thermal particles.
\vspace{.5in}

\noindent{\bf Acknowledgement}

This work was supported in part by the US DOE Grant number DE-FG 02-91ER40685,
by MCT/CNPq as well as by FAPESP, Brazil and by CONICYT, Chile under grant
Fondecyt 1030363 and 7040057 (Int. Coop.).

\appendix

\section{Derivation of Identities for the Largest Time Equation}

In this appendix, we give a brief derivation of the identity used in
(\ref{lte2}) with a generalization to theories with $n$-point
interactions. Let us consider expressions consisting of products of
$n$-factors of the forms 
\begin{eqnarray}
I_{n} & = & \prod_{i=1}^{n} A_{i} - \prod_{i=1}^{n} B_{i},\nonumber\\
\tilde{I}_{n} & = & \prod_{i=1}^{n} A_{i} + \prod_{i=1}^{n} B_{i},
\end{eqnarray}
where $A_{i},B_{i}$ are arbitrary. It follows from this that 
\begin{equation}
\prod_{i=1}^{n} A_{i} = \frac{1}{2}(I_{n}+\tilde{I}_{n}),\quad
\prod_{i=1}^{n} B_{i} = \frac{1}{2}(-I_{n}+\tilde{I}_{n}).
\end{equation}
It follows from this that
\begin{eqnarray}
I_{n+1} & = & A_{n+1}\prod_{i=1}^{n} A_{i} - B_{n+1}\prod_{i=1}^{n}
B_{i}\nonumber\\
& = & \frac{1}{2}(A_{n+1}+B_{n+1}) I_{n} +
\frac{1}{2}(A_{n+1}-B_{n+1})\tilde{I}_{n},\nonumber\\
\tilde{I}_{n+1} & = & A_{n+1}\prod_{i=1}^{n} A_{i} +
B_{n+1}\prod_{i=1}^{n} B_{i}\nonumber\\
& = & \frac{1}{2}(A_{n+1}-B_{n+1})I_{n} + \frac{1}{2}
  (A_{n+1}+B_{n+1})\tilde{I}_{n}.
\end{eqnarray}

These relations can be written in a matrix form as
\begin{equation}
\left(\begin{array}{c}
I_{n+1}\\
\tilde{I}_{n+1}
\end{array}\right) = \frac{1}{2} \left(\begin{array}{cc}
A_{n+1}+B_{n+1} & A_{n+1}-B_{n+1}\\
   &  \\
A_{n+1}-B_{n+1} & A_{n+1}+B_{n+1}
\end{array}\right)\left(\begin{array}{c}
I_{n}\\
\tilde{I}_{n}
\end{array}\right).
\end{equation}
Iterating this, we obtain (we only note the forms of $I_{n}$ which are
relevant for our discussion)
\begin{eqnarray}
I_{1} & = & A_{1}-B_{1},\nonumber\\
I_{2} & = & \frac{1}{2}\left[(A_{1}-B_{1})(A_{2}+B_{2}) +
  (A_{2}-B_{2})(A_{1}+B_{1})\right],\nonumber\\
I_{3} & = & \frac{1}{4}\left[(A_{1}-B_{1})(A_{2}
  +B_{2})(A_{3}+B_{3})\right.\nonumber\\
& & \quad +
  (A_{2}-B_{2})(A_{3}+B_{3})(A_{1}+B_{1})\nonumber\\ 
 & & \quad +(A_{3}-B_{3})(A_{1}+B_{1})(A_{2}+B_{2})\nonumber\\
& & \quad\left.+
   (A_{1}-B_{1})(A_{2}-B_{2})(A_{3}-B_{3})\right],
\end{eqnarray}
and so on. Identifying
\begin{eqnarray}
A_{i} &=& \theta(t-t_{i})\Delta_{aa_{i}}^{(T=0)} (t-t_{i},E_{i}),\nonumber\\
B_{i}&=&\theta(t-t_{i})\Delta_{aa_{i}}^{(T=0)}
    (\underline{t}-t_{i},E_{i}),\label{ab}
\end{eqnarray}
the identity in (\ref{lte2}) follows. If we had a theory with an
$n$-point interaction, we can iterate the above relation $n$ times to
obtain the particular identity. The important thing to note is that in
this difference of products, there will always be a factor of the form
$(A-B)$ in every term and with the identification in (\ref{ab}), such
a factor will always vanish because of (\ref{lte1}). As a result, the
largest time equation holds for any theory.

\section{Spectral Representation for the Propagator at Finite
  Temperature and $\mu\neq 0$} 

In our discussions in this paper, we have worked in a mixed space
representation for the propagator and, as a result, we have been able
to read out the positive and the negative frequency components of the
propagator directly from the definition in  (\ref{frequencydecomp}).  
However, when one works in the momentum space, there is no time
coordinate and, therefore, no direct notion of time ordering. In this
case, the positive and the negative frequency components of a
propagator are obtained from a spectral decomposition of the
propagator. In this appendix, we describe briefly the spectral
decomposition for a scalar propagator at finite temperature in the
presence of a chemical potential. The corresponding analysis for
$\mu=0$ was already discussed in \cite{das:book97} and can be obtained from this
discussion in the limit of a vanishing chemical potential. 

Let us note that since the components of the propagator at finite
temperature depend on both positive and negative frequency components,
one can write a general spectral decomposition for the propagator as 
\begin{eqnarray}
G_{ab}^{(T,\mu)} (x) & = & \int \frac{d^{4}p}{(2\pi)^{4}}\ G_{ab}^{(T,\mu)} (p)\ e^{-ip\cdot x}\nonumber\\
& = &
 i \int_{0}^{\infty} \!\!\!ds \int
 \frac{d^{4}p}{(2\pi)^{4}}\!\left[\frac{\rho_{ab}^{(T,\mu)}(s,p)}{(p_{0}+\mu)^{2} - \vec{p}^{\ 2}-s + i\epsilon}\right.\nonumber\\
 &  &\quad \left. +
   \frac{\tilde{\rho}_{ab}^{(T,\mu)}(s,p)}{(p_{0}+\mu)^{2} -
     \vec{p}^{\ 2}-s - i\epsilon}\right]e^{-ip\cdot x}.\label{spectraldecomp}
\end{eqnarray}
From the structure of the components of the momentum space propagator
in (\ref{mupprop}), we can read out the spectral functions in
(\ref{spectraldecomp}) to be 
\begin{eqnarray}
\rho_{++}^{(T,\mu)} (s, p) & = & (1 + n^{(\mu)} (p_{0}))\delta (s-M^{2}) = -\tilde{\rho}_{--}^{(T,\mu)} (s, p),\nonumber\\
\rho_{+-}^{(T,\mu)} (s, p) & = & \left(\theta(-p_{0}-\mu) + n^{(\mu)}
  (p_{0})\right)\delta (s-M^{2})\nonumber\\
& = &  - \tilde{\rho}_{+-}^{(T,\mu)} (s, p),\nonumber\\
\rho_{-+}^{(T,\mu)} (s, p) & = & \left(\theta(p_{0}+\mu) +
  n^{(\mu)}(p_{0})\right)\delta (s-M^{2})\nonumber\\
& = &  - \tilde{\rho}_{-+}^{(T,\mu)}(s, p),\nonumber\\
\rho_{--}^{(T,\mu)} (s, p) & = & n^{(\mu)} (p_{0}) \delta (s-M^{2}) = - \tilde{\rho}_{++}^{(T,\mu)} (s, p).\label{spectralfunction}
\end{eqnarray}
We note that when $\mu=0$, these reduce to the spectral functions
already described in \cite{das:book97}. For $T=0$, the spectral functions are given
by 
\begin{eqnarray}
\rho_{++}^{(T=0,\mu)} (s, p) & = & \delta (s-M^{2}) = - \tilde{\rho}_{--}^{(T=0,\mu)} (s, p),\nonumber\\
\rho_{+-}^{(T=0,\mu)} (s, p) & = & \theta (-p_{0}-\mu)\delta (s-M^{2}) = - \tilde{\rho}_{+-}^{(T=0,\mu)} (s, p),\nonumber\\
\rho_{-+}^{(T=0,\mu)} (s, p) & = & \theta (p_{0}+\mu)\delta (s-M^{2}) = - \tilde{\rho}_{-+}^{(T=0,\mu)} (s, p),\nonumber\\
\rho_{--}^{(T=0,\mu)} (s, p) & = & 0 = \tilde{\rho}_{++}^{(T=0,\mu)} (s, p).
\end{eqnarray}

The components of the propagator are time ordered and carrying out the
integration in (\ref{spectraldecomp}) over the $p_{0}$ variable would,
in fact, show this and lead to the frequency decomposition 
\begin{equation}
G_{ab}^{(T,\mu)} (x)  =  \theta(x^{0}) G_{ab}^{(T,\mu)(+)} (x) + \theta (-x^{0}) G_{ab}^{(T,\mu)(-)} (x).
\end{equation}
In fact, comparing with (\ref{spectraldecomp}), we can now determine
\begin{eqnarray}
\lefteqn{G_{ab}^{(T,\mu) (\pm)} (x)}\nonumber\\
 & = & \int_{0}^{\infty} ds\int \frac{d^{4}p}{(2\pi)^{4}}\ (2\pi)\delta ((p_{0}+\mu)^{2}-\vec{p}^{\ 2}-s)\ e^{-ip\cdot x}\nonumber\\
 &\times&\left[\theta (\pm (p_{0}+\mu)) \rho_{ab}^{(T,\mu)} (s, p) -
   \theta (\mp (p_{0}+\mu)) \tilde{\rho}_{ab}^{(T,\mu)} (s,
   p)\right].\nonumber\\
& & 
 \end{eqnarray}
 The positive and the negative frequency components in the mixed space
 can be related to the spectral functions as 
\begin{eqnarray}
\lefteqn{G_{ab}^{(T,\mu) \pm} (t, E)}\nonumber\\
 & = & \int_{0}^{\infty} ds \int dp_{0}\ \delta ((p_{0}+\mu)^{2} - E^{2} - s + M^{2}) e^{-ip_{0}t}\nonumber\\
  &\times& \left[\theta (\pm (p_{0}+\mu)) \rho_{ab}^{(T,\mu)} (s, p) -
    \theta (\mp (p_{0}+\mu)) \tilde{\rho}_{ab}^{(T,\mu)} (s,
    p)\right].\nonumber\\
& & 
  \end{eqnarray}
  Recalling that all the spectral functions in
  (\ref{spectralfunction}) are proportional to $\delta (s-M^{2})$
  and that the remaining factor depends only on $p_{0}$, let us define 
\begin{eqnarray}
  \rho_{ab}^{(T,\mu)} (s,p) &=& \delta(s-M^{2}) \rho_{ab}^{(T,\mu)}
  (p_{0}),\nonumber\\
 \tilde{\rho}_{ab}^{(T,\mu)}(s, p) &=& \delta (s-M^{2}) \tilde{\rho}_{ab}^{(T,\mu)} (p_{0}).
  \end{eqnarray}
  Then, the positive and the negative frequency components in the
  mixed space can also be written as (doing the $s$ integration) 
\begin{eqnarray}
\lefteqn{G_{ab}^{(T,\mu) \pm} (t,E)
 = \int dp_{0} \delta ((p_{0}+\mu)^{2}-E^{2})\ e^{-ip_{0}t}}\nonumber\\
   & & \times\left[\theta(\pm (p_{0}+\mu)) \rho_{ab}^{(T,\mu)} (p_{0})
     - \theta (\mp (p_{0}+\mu)) \tilde{\rho}_{ab}^{(T,\mu)}
     (p_{0})\right],\nonumber\\
& & 
  \end{eqnarray}
which are the expressions  used in the text (although we have derived
these directly from the expressions of the components of the
propagators in the mixed space).

\end{document}